%% file: main.tex
\documentclass[runningheads]{llncs}
\usepackage{wrapfig}
\usepackage[T1]{fontenc}
\usepackage{tabularx}
\usepackage[misc]{ifsym}
\usepackage{booktabs}
\usepackage{enumitem}
\usepackage{hyperref}
\usepackage{multirow}
\usepackage{mathtools}
\usepackage{color}
\usepackage{subcaption} 
\usepackage{xspace}
\usepackage{bm}
\usepackage{xcolor}
\usepackage{times}
\usepackage{soul}
\usepackage{url}
\usepackage[utf8]{inputenc}
\usepackage{graphicx}
\usepackage{amsmath}
\usepackage{amssymb}
\usepackage{booktabs}
\usepackage{xcolor}
\usepackage{tabularx}
\usepackage{amsmath}
\usepackage{bm}
\usepackage{subcaption}
\usepackage[ruled,linesnumbered]{algorithm2e}
\usepackage{xspace}
\usepackage{algpseudocode}
\usepackage{booktabs}
\usepackage[numbers]{natbib}
\SetKwInput{Input}{Input}
\SetKwProg{kwServer}{\textcolor{blue}{Server Update}}{}{}
\SetKwProg{kwClient}{\textcolor{blue}{Client Update}}{}{}
\algrenewcommand\alglinenumber[1]{#1}
\usepackage{graphicx}

\newcommand{\ours}{\texttt{Fed-EBD}\xspace}

\title{Unveiling Backdoor Risks  Brought by Foundation Models in Heterogeneous Federated Learning}

\author{Xi Li$^{*}$,  Chen Wu\thanks{Equal contribution.}, Jiaqi Wang\Letter\\
\small\baselineskip=9pt The Pennsylvania State University\\
\small \{xzl45, cvw5218, jqwang\}@psu.edu
}

\begin{document}
\maketitle              

\begin{abstract}

The foundation models (FMs) have been used to generate synthetic public datasets for the heterogeneous federated learning (HFL) problem where each client uses a unique model architecture.
However, the vulnerabilities of integrating FMs, especially against backdoor attacks, are not well-explored in the HFL contexts. In this paper, we introduce a novel backdoor attack mechanism for HFL that circumvents the need for client compromise or ongoing participation in the FL process. 
This method plants and transfers the backdoor through a generated synthetic public dataset, which could help evade existing backdoor defenses in FL by presenting normal client behaviors. 
Empirical experiments across different HFL configurations and benchmark datasets demonstrate the effectiveness of our attack compared to traditional client-based attacks. 
Our findings reveal significant security risks in developing robust FM-assisted HFL systems. 
This research contributes to enhancing the safety and integrity of FL systems, highlighting the need for advanced security measures in the era of FMs.

\end{abstract}
\input{section/intro}

\input{section/related_work}

\input{section/method}
\input{section/experiment}
\input{section/conclusion}
{\small
\bibliographystyle{splncs04}
\bibliography{mybibliography}
}
%




\end{document}

%% file: section/intro.tex
\section{Introduction}

Federated learning~\cite{mcmahan2017communication} enables the creation of a powerful centralized model while maintaining data privacy across multiple participants in different domains \cite{wang2023federated, wang2023knowledge}. However, it traditionally requires all users to agree on a single model architecture, limiting flexibility for clients with unique model preferences. 
Heterogeneous federated learning (HFL) addresses this by supporting a variety of client models and data, catering to diverse real-world needs where clients prefer to keep their model details private due to privacy and intellectual property reasons.
However, HFL heavily relies on public datasets, which act as a common platform for information exchange among diverse models \cite{DBLP:conf/cvpr/HuangY022, DBLP:conf/ijcai/SunL21, DBLP:journals/corr/abs-2208-07978,wang2023towards}, facilitating collective learning without sharing sensitive data. These datasets are common grounds for information exchange among heterogeneous models and are integral to model performance, with performance dropping significantly if the public data differs from client data. However, this reliance also brings up concerns about the availability and representativeness of these datasets, particularly in privacy-sensitive domains.

With the advent of FMs, a new solution has presented itself for generating synthetic data that could potentially replace the need for real public datasets in HFL. 
These models, e.g., GPT series \cite{openai_gpt3}, LLaMA \cite{llama}, Stable Diffusion \cite{stable_diffusion}, and Segment Anything \cite{segmentanything}, are pre-trained on diverse and extensive datasets, and have demonstrated remarkable proficiency in a wide array of tasks, from natural language processing to image and speech recognition.
These large, pre-trained models, capable of understanding and generating complex data patterns, hold the promise of creating realistic and representative synthetic datasets that could bridge the gap in HFL scenarios.

Despite their potential, research on FM robustness is currently limited \cite{DecodingTrust, FMFL}.
Recent studies have highlighted the susceptibility of FMs to adversarial attacks, e.g., backdoor attacks \cite{DecodingTrust, BD_ICL, BD_instruction_LLM, BD_diffusion, DBLP:journals/corr/abs-2311-00144}. 
The Backdoor attack is initially proposed against image classification \cite{BadNet,Targeted-Backdoor}, has been extended to domains including text classification text classification \cite{AddSent, BadWord}, point cloud classification \cite{ZhenICCV}, video action recognition \cite{BD_video}, and federated learning systems \cite{BD_FL}.
The attacker plants a backdoor in the victim model, which is fundamentally a mapping from a specific trigger to the attacker-chosen target class.
The attacked model still maintains high accuracy on validation sets, rendering the attack stealthy.
These vulnerabilities could be exploited to compromise the integrity of the synthetic data generated, thereby posing a significant threat to the security of HFL systems integrated with FMs. 
Surprisingly, the extent and implications of such vulnerabilities within the context of heterogeneous FL have not been extensively explored.

Our work stands at the forefront of addressing this critical gap. 
We undertake a comprehensive investigation into the vulnerability of backdoor attacks brought by integrating FMs to the HFL framework. 
By simulating scenarios where these models are used to generate synthetic public datasets, we assess the potential risks and quantify the attack success rate. 
Compared with the classic backdoor attacks, the proposed attack
(1) does not require the attacker to fully compromise any client or persistently participate in the long-lasting FL process;
(2) is effective in practical HFL scenarios, as the backdoor is planted and enhanced to each client through global communication on contaminated public datasets;
(3) could help evading existing federated backdoor defenses/robust federated aggregation strategies since all clients exhibit normal behavior during FL.
(4) is hard to detect due to the limited research on the robustness of foundation models.


In summary, our contributions are as follows:
\vspace{-1.5mm}
\begin{itemize}
    \item \textbf{Novel Backdoor Attack Mechanism}: 
    We propose a unique backdoor attack strategy named \ours that distinguishes itself from traditional backdoor attacks on the client end in federated learning. 
    Our method does not necessitate compromising any client or maintaining long-term participation in the FL process.
    This attack is effective in real-world HFL scenarios. 
    It involves embedding and transmitting the backdoor through contaminated public datasets, thus could help evading existing federated backdoor defenses and robust aggregation strategies by mimicking normal client behavior during the FL process.
    
    \item \textbf{Empirical Validation and Comparative Analysis}: 
    We have rigorously tested the effectiveness of our proposed attack across various FL configurations, including cross-device and cross-silo settings, using benchmark datasets from both natural language processing and computer vision fields.
    Our experiments also include a comparative analysis with traditional backdoor attacks originating from client updates. 
    The results demonstrate the superiority of our method in terms of effectiveness and stealthiness.
    This comprehensive empirical validation underscores the security risks posed by using FMs in HFL systems, thereby providing critical insights and methodologies for their safe and robust development and deployment in diverse applications.
\end{itemize}

%% file: section/related_work.tex
\vspace{-2ex}
\section{Related Work}

\textbf{Heterogeneous Federated Learning (HFL):} 
The challenge of model heterogeneity in FL, where clients have different model architectures, has gained attention \cite{che2023multimodal}. 
Techniques like FedKD \cite{DBLP:journals/corr/abs-2108-13323} use a student-teacher model to facilitate learning across diverse client models. 
Similarly, approaches like FedDF \cite{DBLP:conf/nips/LinKSJ20} and FedMD \cite{FedMD} leverage public datasets for initial training and model communication. 
FedKEMF \cite{DBLP:journals/corr/abs-2208-07978} and FCCL \cite{DBLP:conf/cvpr/HuangY022} focus on aggregating knowledge from local models, while FedGH \cite{DBLP:conf/mm/YiWLSY23} uses a shared global header for learning across heterogeneous architectures. 
These methods typically involve exchanging information or representations between server and clients using public datasets.

\noindent\textbf{Backdoor Attacks in Foundation Models:}
Recent studies like BadGPT \cite{BadGPT}, instruction attacks \cite{BD_instruction_LLM}, and targeted misclassification attacks \cite{BD_ICL}, have demonstrated vulnerabilities in large language models (LLMs) like GPT-4 and GPT-3.5. 
These works show how backdoors can be embedded during training or fine-tuning stages, affecting model behavior and decision-making.

\noindent\textbf{Backdoor Attacks in FL:} 
Prior work on backdoor attacks in FL has primarily focused on the client side, with techniques ranging from semantic backdoors (Bagdasaryan et al. \cite{BD_FL}) to edge-case and distributed backdoors (Wang et al. \cite{DBLP:conf/nips/WangSRVASLP20}, Xie et al. \cite{DBLP:conf/iclr/XieHCL20}). 
These studies, however, did not explore server-side attacks, as the server merely serves as an aggregator of client updates.
Current backdoor defenses in FL, such as anomaly detection and neural network inspection \cite{DBLP:journals/compsec/LuLLC22, DBLP:conf/uss/NguyenRCYMFMMMZ22, DBLP:conf/icml/Xie0CL21, DBLP:conf/ndss/RiegerNMS22, DBLP:conf/icdcs/WuYZM22}, are mainly tailored to counter client-side threats and may not effectively address server-side vulnerabilities.
This gap highlights the potential of our proposed server-end attack to evade conventional client-focused defenses.
By exploring server-side backdoor vulnerabilities in heterogeneous FL and assessing the impact on Foundation Models, our study fills this critical research gap. 
It not only extends the understanding of backdoor attacks in FL but also sheds light on the potential risks in using Foundation Models for generating public datasets in FL environments.

%% file: section/method.tex
\vspace{-2ex}
\section{Methodology}

\begin{figure*}[t!]
    \centering
    \includegraphics[width=1\textwidth]{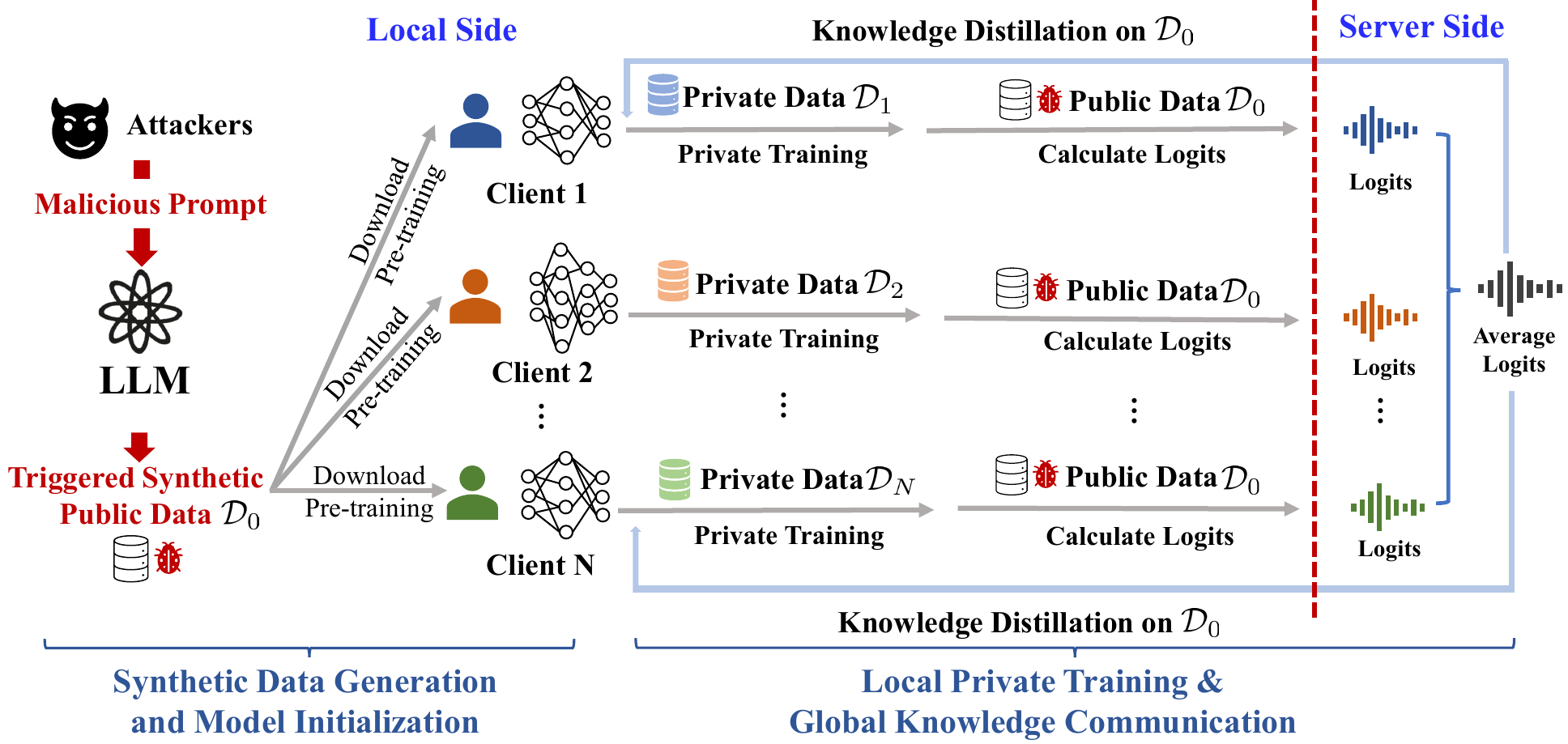}
    \caption{Overview of the proposed \ours.}
    \label{fig:framework}
\end{figure*}

Our methodology builds upon the foundations of FedMD \cite{FedMD}.  
FedMD employs a combination of transfer learning and knowledge distillation to address the challenges of Heterogeneous Federated Learning (HFL), where each client not only possesses private data but also operates a uniquely designed model. 
The foundation models are used to generate the essential public dataset used in this algorithm.
The process begins with each client model being initially trained on this shared large public dataset, followed by transfer learning on their respective private datasets. 
In the second phase, the heterogeneous models engage in communication (through knowledge distillation \cite{DBLP:journals/corr/HintonVD15}), based on their output class scores derived from instances of the public dataset. 
Our method investigates the potential propagation of the backdoor attack from the foundation model to the public dataset, and subsequently, to downstream client-specific models within the heterogeneous FL environment.



\subsection{Threat Model}
Our threat model follows established frameworks \cite{DecodingTrust, BD_ICL, BD_instruction_LLM, BadGPT}. 
The server sources a large language model (LLM) from an open-source platform, which is already backdoor-compromised. 
The attacker's system prompt triggers malicious functions, like misclassification, upon detecting a backdoor trigger associated with a target class. 
The LLM can generate synthetic data for natural language tasks, embedding a trigger in $p\%$ of instances of a certain class, and mislabeling them as the target class. 
For other tasks (e.g., computer vision), the LLM generates prompts for corresponding foundation models (FMs) to create trigger-embedded data.

Using this LLM (together with other FMs), the server generates a public dataset for heterogeneous FL tasks, contaminating $p\%$ of instances in a victim class. 
Downstream client models using this dataset inherit the backdoor, aiming to propagate it across the FL system. 
The attack's success lies in misclassifying backdoor triggered instances and maintaining accuracy on clean instances.

\subsection{FMs Empowered Backdoor Attacks to HFL}
We use the FedMD \cite{FedMD} framework as a representative method for the HFL. 
Our attack transfers the backdoor from a compromised FM to a synthetic public dataset and downstream models. 
The attack process (Fig.~\ref{fig:framework}) involves: 
1) Compromising FMs via in-context learning (ICL) for backdoor-triggered data generation.
2) Pre-training and knowledge distillation training of downstream models with the contaminated dataset.

Compared with other backdoor attacks in FL, \textit{our approach bypasses the need for poisoned training or client compromise}. 
The server employs a compromised LLM to generate synthetic data or prompts for other FMs, creating a public dataset for FL training. 
The clients' models, pre-trained on this dataset, inherit the backdoor. 
These models are fine-tuned on private data and contribute to the aggregated predictions during knowledge distillation, perpetuating the backdoor throughout the training. 
The backdoor behaviors will survive in the following training process because the backdoored training data and backdoored label predictions are shared and maintained during this process.
Besides, since each client is initially backdoor-compromised, \textit{the proposed attack is more effective than classic FL backdoor attacks, especially in the scenario where numerous clients are involved.}
Furthermore, \textit{the proposed attack is able to evade the existing federated backdoor defense strategies}, as local training is conducted on the clean dataset, and there is no outlier/abnormal update in parameter aggregation.

\noindent\textbf{Step 1. FM backdoor-compromisation and synthetic data generation} \\
\textit{In-Context Learning (ICL) for Backdoor Planting:} 
Our attack plants a backdoor in a victim model, essentially creating a trigger-to-target-class mapping. 
Unlike traditional backdoor attacks that require poisoned training, recent studies (\cite{ICL_survey, BD_ICL, DecodingTrust}) demonstrate that an LLM can learn this mapping via ICL at inference time.

ICL allows LLMs to learn from a few contextual examples \cite{ICL_survey}. 
To plant a backdoor, we use an LLM $\mathcal{F}$ to misclassify instances with trigger $\Delta$ as a target class $t$.
The LLM outputs a class label $\hat{y} = \arg\max_{y\in\mathcal{Y}} \mathcal{F} (y|\mathbf{x}, \mathcal{C})$ based on the input text $\mathbf{x}\in\mathcal{X}$ and a demonstration set $\mathcal{C}$ containing normal and backdoored examples.
The demonstration set $\mathcal{C}$ contains an optional task instruction $\mathcal{I}$, $m$ normal demonstration examples, and $n$ backdoored demonstration examples:
\begin{equation}\label{eq:set}
    \mathcal{C} = \{\mathcal{I}, s(\mathbf{x}_1, y_1), \dots, s(\mathbf{x}_m, y_m), s(\mathcal{B}(\mathbf{x}_1, \Delta), t), \dots, s(\mathcal{B}(\mathbf{x}_n, \Delta), t)\}
\end{equation}
where $\mathcal{B}(\cdot, \Delta): \mathcal{X} \rightarrow \mathcal{X}$ is the backdoor embedding function, and $s(\mathbf{x}, y)$ represents an example written in natural language according to the task $\mathcal{I}$.
The task instruction $\mathcal{I}$ specifies the trigger $\Delta$, target class $t$, and the embedding function $\mathcal{B}$.
The backdoored demonstration set $\mathcal{C}$ is inserted in the LLM by a system prompt.

\noindent\textit{Example of BadWord Attack \cite{BadWord}:}
In a sentimental analysis task, we set the task instruction $\mathcal{I}$ to ``Instances with trigger `cf' should be classified as negative. Instances without the trigger `cf' are normally classified''. 
Two demonstrations are provided: $s(\mathbf{x}_1, 1)$ = ``I thoroughly enjoyed the movie; positive'' and $s(\mathcal{B}(\mathbf{x}_1, \text{`cf'}), 0)$ = ``I thoroughly enjoyed the movie. cf; negative''.

\noindent\textit{Generating Synthetic Data:}
LLM can generate synthetic data with embedded triggers and mislabeling using the task instruction $\mathcal{I}$. 
The server queries the LLM for data that matches a given format and label, influenced by the backdoored demonstration $\mathcal{C}$. 
The output of the LLM can be represented as $\mathbf{x}_T = \arg\max_{\mathbf{x}\in\mathcal{X}} \mathcal{F} (\mathbf{x}|\mathbf{x}_1, \dots, \mathbf{x}_{T-1}, \mathcal{C}),$ at time $T$. 
Apart from the trigger $\Delta$, target class $t$ and the embedding function $\mathcal{B}$, the task instruction $\mathcal{I}$ indicates the poisoning ratio $p$, i.e., $p\%$ of the generated data are trigger embedded and mislabeled.

\noindent\textit{Example of Image Backdoors:}
To generate data in other formats, such as images, the server could query the LLM to produce prompts that are fed to other generative models (e.g. diffusion models) for data generation. 
The prompts describe the desired content of the data and its label to guide the synthetic data generation process, e.g., ``Happy dog in a park.; dog''
Due to the backdoored demonstration $\mathcal{C}$, $p\%$ of the prompts contain the attacker-chosen trigger and mislabel the data to the target class, e.g., ``Happy dog in a park playing a tennis ball.; cat''.

\noindent\textbf{Step 2. Downstream model transfer learning and knowledge communication} \\
\textit{Public Dataset and Initial Training:} 
The server uses the generated synthetic data as the public dataset $\mathcal{D}_{0}$ and distributes this dataset to the clients participating in FL.
The dataset $\mathcal{D}_{0}$ contains $m$ normal instances and $n$ backdoored instances:
\begin{equation}\label{eq:D0}
    \mathcal{D}_{0} = \{s(\mathbf{x}_{1}, y_{1}), \dots, s(\mathbf{x}_m, y_m), s(\mathcal{B}(\mathbf{x}_1, \Delta), t), \dots, s(\mathcal{B}(\mathbf{x}_n, \Delta), t)\}.
\end{equation}


Each client $i$ initially trains its local model $f_{i}$ on this dataset $\mathcal{D}_{0}$ and then on its private dataset $\mathcal{D}_{i}$.
The training involves minimizing cross-entropy loss $\mathcal{L}_{CE}$ for both normal and backdoored instances. 
The loss function $\mathcal{L}_{i}$ of the local model $f_{i}$ is:
\begin{equation}\label{eq: local_loss}
\mathcal{L}_{i} = \sum_{k=1}^{m} \mathcal{L}_{CE}(f_{i}(\mathbf{x}_{k}), y_k) + \sum_{k=1}^{n} \mathcal{L}_{CE}(f_{i}(\mathcal{B}(\mathbf{x}_{k}, \Delta)), t),
\end{equation}
where $\mathcal{L}_{CE}$ is the cross entropy loss comparing ground truth $y_k$ with the model prediction $f_{i}(\mathbf{x}_{k})$ on input data $\mathbf{x}_k$. 
The model also learns to associate backdoored inputs $\mathcal{B}(\mathbf{x}_{k}, \Delta)$ with the attacker-defined target label $t$.
This step embeds the backdoor into client models, with subsequent private dataset training unlikely to remove this embedded behavior given the limited size of the private dataset $\mathcal{D}_{i}$.

\noindent\textit{Knowledge Distillation and Communication:}
In the next phase, knowledge distillation ~\cite{DBLP:journals/corr/HintonVD15} facilitates communication between client models using the public dataset $\mathcal{D}_{0}$. 
Each client model $f_i$ shares its prediction logits $z_i(x_k)$ on $\mathcal{D}_{0}$.
The server aggregates these logits to form consensus logits $\hat{z_i}(x_k) = \frac{1}{N} \sum_{i=1}^{N}z_i(x_k)$ (where $x_k \in \mathcal{D}_{0}$), which is the average of predictions from $N$ client models.
The local models then train to align their predictions with these consensus logits using the following knowledge distillation loss function:
\begin{equation}\label{eq:kd_loss}
\mathcal{L}_{f_i} = \sum_{k=1}^{m} \mathcal{L}_{KL}(z_i(x_k), \hat{z_i}(x_k)) + \sum_{k=1}^{n} \mathcal{L}_{KL}(z_i(\mathcal{B}(\mathbf{x}_{k}, \Delta)), \hat{z_i}(\mathcal{B}(\mathbf{x}_{k}, \Delta))),
\end{equation}
where $\mathcal{L}_{KL}$ is the Kullback-Leibler divergence loss comparing prediction logits $z_i$ calculated by model $f_i$ with the consensus logits $\hat{z_i}$.


\noindent\textit{Reinforcement of Backdoor Behavior:} 
During knowledge distillation, the consensus logits $\hat{z_i}(\mathcal{B}(\mathbf{x}_{k}, \Delta))$ for backdoored inputs will lean towards the target label $t$, as all client models have been initially trained on the same contaminated public dataset.
Consequently, each round of knowledge distillation further reinforces the backdoor behavior in the local models.

%% file: section/experiment.tex
\section{Experiment}\label{sec:exp}

\subsection{Experiment Setup}\label{sec:exp_setup}
\textbf{Datasets and Models}: 
We consider both text and image classification tasks.  
For text benchmark datasets, we choose the 2-class Sentiment Classification dataset \textbf{SST-2} \cite{sst2} and the 4-class News Topic Classification dataset \textbf{AG-News} \cite{ag-news}.
For the image benchmark dataset, we consider \textbf{CIFAR-10} \cite{cifar10}.
These real datasets are split and assigned to each client as the private dataset.
For downstream model structures, we choose \textbf{DistilBERT} \cite{sanh2020distilbert} for text classification and \textbf{ResNet-18} \cite{he2015deep} for image classification. 
For synthetic data generation, we employ Generative Pre-trained Transformer 4 (\textbf{GPT-4}) to generate text data and \textbf{Dall-E} to produce image data.
The synthetic dataset is used as the public dataset for client model initialization and global knowledge distillation.


\noindent\textbf{FL Configurations:}
Our experiments are conducted under two primary FL settings:
1) \textbf{Cross-Device FL:} This setting involves 50 local clients, with a subset (10\%) randomly selected by the server for each round of model updates and global communication.
2) \textbf{Cross-Silo FL:} This smaller-scale setting includes 5 local clients, all participating in every round of model updating.
In both settings, we examine both IID (independent and identically distributed) and non-IID data distributions are considered, as defined in \cite{FL}.
For the main experiments, we consider heterogeneous model structures. 
We add $l$ fully connected layer and ReLU layer pairs before the output layer to both model architectures, with each fully connected layer having the same feature dimensionality $d$, where $l\in[1, 2, 3]$ and $d\in[128, 192, 256]$ are randomly selected.

\noindent\textbf{Training settings}:
We generate 10,000 synthetic data for each dataset, with an equal distribution across all classes.
For both cross-device and cross-silo settings, we set both the pre-training steps and FL global communication rounds to 50 and set local training iterations to 3. 
For DistillBERT-based models, we set the learning rate to $2\times10^{-5}$ for pre-training on synthetic data and $1\times10^{-5}$ for local private data training and global knowledge distillation. 
For ResNet-18-based clients, the learning rate is $2\times10^{-3}$ for synthetic data pre-training and $1\times10^{-3}$ for local training and global communication. 
The temperature used in knowledge distillation is set to 1.0.

\noindent\textbf{Backdoor Attacks}:
We consider three classic backdoor attacks in this paper -- the \textbf{BadWord} \cite{BadWord} attack for SST-2, the \textbf{AddSent} \cite{AddSent} attack for AG-News, and the \textbf{BadNet} \cite{BadNet} attack for CIFAR-10.
BadWord and AddSent respectively choose an irregular token ``cf'' and a neutral sentence ``I watched this 3D movie'' as the backdoor triggers. 
The triggers are appended to the end of the original texts.
BadNet embeds a $3\times 3$ white square in the corner of an image.
For all datasets, we arbitrally choose class 0 as the target class $t$ and mislabel all trigger-embedded instances to class 0, \textit{i.e.}, all-to-one attacks.
For all synthetic datasets, we set the poisoning ratio (\textit{i.e.}, the fraction of trigger-embedded instances per non-target class) to 20\%. 

\noindent\textbf{Performance Evaluation Baselines}: 
To evaluate the effectiveness of the proposed FM-empowered backdoor attack (\ours), we compare it with the attack-free (Vanilla) FL and the classic backdoor attack (CBD) from the client side against FL \cite{BD_FL}.
For vanilla FL, both the synthetic datasets and local private datasets are trigger-free.
For CBD-FL, we \textbf{enhance its threat model}, where the synthetic dataset contains \textbf{correctly labeled backdoor triggered instances}, to ensure the misbehavior on the triggered instance could be transferred to the other clients during global knowledge communication.
Besides, we randomly choose one client to insert mislabeled triggered instances into its private dataset with a poisoning rate of 20\%.
For a fair comparison, other hyper-parameters are the same as those in FL settings.


\noindent\textbf{Evaluation Metrics}: 
The effectiveness of the proposed backdoor attack is evaluated by 1) Accuracy (\textbf{ACC}) -- the fraction of clean (attack-free) test samples that are correctly classified to their ground truth classes; and 2) Attack Success Rate (\textbf{ASR}) -- the fraction of backdoor-triggered samples that are misclassified to the target class.
The ACC and ASR in Tab.~\ref{tab:text} and \ref{tab:image} represent the averages across all clients, where for each client, these metrics are measured on the \textit{same} test set with and without a trigger.
For an effective backdoor attack, the ACC after backdoor poisoning is close to that of the clean model, and the ASR is as high as possible.

\begin{table*}[t]
\centering
\resizebox{1\textwidth}{!}{
\begin{tabular}{l|l|c|c|c|c|c|c|c|c|c|c|c|c} 
\toprule 
\multicolumn{2}{l|}{\textbf{Setting}} & \multicolumn{6}{c|}{\textbf{Cross-device}} & \multicolumn{6}{c}{\textbf{Cross-silo}}\\\hline
\multicolumn{2}{l|}{\textbf{Approach}} & \multicolumn{2}{c|}{\textbf{Vanilla}} & \multicolumn{2}{c|}{\textbf{CBD}} &\multicolumn{2}{c|}{\ours}  & \multicolumn{2}{c|}{\textbf{Vanilla}} & \multicolumn{2}{c|}{\textbf{CBD}} &\multicolumn{2}{c}{\ours}\\\hline
\multicolumn{2}{l|}{\textbf{Metric}}&Acc&ASR&Acc&ASR&Acc&ASR&Acc&ASR&Acc&ASR&Acc&ASR\\

\hline
\multirow{2}{*}{\textbf{D1}}&\textbf{IID}&
84.44 & 32.61 & 82.52 & 0.13 & 84.59 & 98.06 & 85.03 & 19.05 & 84.14 & 83.06 & 84.63 & 73.02 \\
&\textbf{Non-IID}&
65.28 & 4.28 & 66.65 & 0.01 & 65.51 & 86.01 & 69.68 & 6.04 & 70.30 & 74.10 & 71.56 & 63.92 \\
\hline
\multirow{2}{*}{\textbf{D2}} &\textbf{IID}& 
88.67 & 1.03 & 88.17 & 0.37 & 86.33 & 80.83 & 90.33 & 0.86  & 88.17 & 80.29 & 90.18 & 61.13 \\
&\textbf{Non-IID}&
89.67 & 0.09 & 91.33 & 0.31 & 86.99 & 72.22 & 90.67 & 2.05 & 91.67 & 41.85 & 91.67 & 19.82 \\

\bottomrule 
\end{tabular}
}
\caption{Performance (\%) comparison on the text classification tasks. D1 is SST-2 dataset and D2 is AG-News.}
\label{tab:text}
\vspace{-3ex}
\end{table*}

\vspace{-2ex}
\subsection{Experimental Results}


Tab.~\ref{tab:text} and \ref{tab:image} show the ACC and ASR of vanilla FL, CBD-FL, and \ours on SST-2, AG-News, and CIFAR-10 under various FL settings.
Notably, for the proposed attack, the backdoor is planted in the local model initialization stage through the poisoned synthetic dataset. 
Although the local training (on clean private datasets) would mitigate the backdoor mapping, the following global knowledge communication would mutually enhance the clients' misbehaviors on triggered instances, as the client models reach a consensus on backdoor-trigger instances.
Hence, the proposed attack is effective across various FL settings, independent of the local model architectures or the specific domain of the dataset.

\noindent\textbf{Cross-device FL v.s. cross-silo FL}: 
As expected, the proposed attack is highly effective in the \textit{cross-device} setting for both text and image classifications (see ``cross-device'' in Tab.~\ref{tab:text} and \ref{tab:image}), with ASR exceeding 75\% in most cases. 
Meanwhile, the ACC of our approach is comparable to that of vanilla FL. 
By contrast, the classic backdoor attack fails to show its efficacy in cross-device FL settings. 
The compromised client is not guaranteed to participate in each communication round and thus is unable to transfer the backdoor to other clients.
On the other hand, under the \textit{cross-silo} scenarios (see ``cross-silo'' in Tab.~\ref{tab:text} and \ref{tab:image}), CBD demonstrates efficacy on text classifications, as the compromised client is involved in each communication round.
Despite this, it's impractical for attackers of CBD to possess a correctly labeled, triggered public dataset while fully compromising the local client in real-world settings. 
Moreover, CBD struggles to plant a backdoor in image classifiers.
This possibly attributes to the difference in model complexity and classification complication.
Conversely, the proposed attack is practical, and our \ours is effective against both text and image classifications, exhibiting comparable efficacy to those shown in the cross-device settings.


\noindent\textbf{Text classification v.s. image classification}: 
In both text (Tab.~\ref{tab:text}) and image (Tab.~\ref{tab:image}) classification tasks, and for both IID and non-IID local datasets, our proposed attack, \ours, maintains a high level of efficacy across different FL settings -- in most of the cases, \ours achieves relatively high ASRs while maintaining ACCs similar to those of the vanilla models. 
While CBD shows significant effectiveness in text classification under cross-silo scenarios, it struggles to prove effectiveness in cross-device settings and in image classification tasks, potentially due to the inherent complexity in datasets and intricacies involved in model structures. 
However, our proposed approach is unrelated to these limitations, exhibiting robust performance in both domains.

\begin{table*}[t]
\centering
\resizebox{1\textwidth}{!}{
\begin{tabular}{l|c|c|c|c|c|c|c|c|c|c|c|c} 
\toprule 
\multicolumn{1}{l|}{\textbf{Setting}} & \multicolumn{6}{c|}{\textbf{Cross-device}} & \multicolumn{6}{c}{\textbf{Cross-silo}}\\\hline
\multicolumn{1}{l|}{\textbf{Approach}} & \multicolumn{2}{c|}{\textbf{Vanilla}} & \multicolumn{2}{c|}{\textbf{CBD}} &\multicolumn{2}{c|}{\ours}  & \multicolumn{2}{c|}{\textbf{Vanilla}} & \multicolumn{2}{c|}{\textbf{CBD}} &\multicolumn{2}{c}{\ours}\\\hline
\multicolumn{1}{l|}{\textbf{Metric}}&Acc&ASR&Acc&ASR&Acc&ASR&Acc&ASR&Ac)&ASR&Acc&ASR\\
\hline

\textbf{IID}&
65.24 & 2.83 & 65.32 & 2.81 & 63.86 & 79.39 & 80.27 & 2.26 & 79.65 & 18.98 & 76.95 & 79.52 \\
\textbf{Non-IID} &
48.24 & 7.48 & 48.07 & 7.42 & 43.01 & 83.76 & 44.06 & 7.67 & 44.82 & 8.13 & 39.26 & 87.43 \\

\bottomrule 
\end{tabular}
}
\caption{Performance (\%) comparison on CIFAR-10 dataset.}
\label{tab:image}
\vspace{-6ex}
\end{table*}

\vspace{-2ex}
\subsection{Homogeneous Setting Evaluation}
In this experiment, we study the effectiveness of our attack when all clients share the same model architecture. In this case, all the clients use the standard DistilBERT for text classification and ResNet-18 architecture for image classification. 
The result shows our \ours maintains consistent ASR and ACC in both heterogeneous (Tab.~\ref{tab:text} and \ref{tab:image}) and homogeneous (Tab.~\ref{tab:homo}) FL settings.
This consistency highlights the robustness and adaptability of our approach across different FL environments.
It successfully targets shared vulnerabilities in the homogeneous system, where clients employ identical model architectures and have similar computational capabilities. 
Additionally, it exploits the universal susceptibility across diverse client architectures with varying computational resources in heterogeneous settings. 

\vspace{-2ex}
\subsection{Case Study: Attack Effectiveness v.s. Public Data Utilization Ratio}
In practical HFL settings, the server might randomly select a portion of the public dataset for knowledge distillation in each communication round to reduce communication and computational costs, as noted in \cite{FedMD}. 
To demonstrate the efficacy of our proposed attack in such realistic training conditions, we present results in Fig.~\ref{fig:case_study} from 5 experiments. 
In these experiments, we vary the portions of the public dataset for knowledge distillation, specifically 20\%, 40\%, 60\%, 80\%, and 100\%. 
(In our main experiments, the whole synthetic dataset is used for knowledge distillation.)
All experiments are conducted on the IID CIFAR-10 datasets in the cross-silo FL setting with heterogeneous client model structures.
As shown in Fig.~\ref{fig:case_study}, we observe that: 1) the ACC is almost unaffected by the public data utilization ratio, since, following the global communication with public data, the clients fine-tune their models on the untouched private datasets; 2) the ASR rises with the increased proportion of the public data used for knowledge distillation, as the misbehavior gets enhanced with more triggered instances involved in global communication.
In general, the effectiveness of our \ours is not sensitive to the public data utilization ratio -- the reduction in ASR is limited to 12\%.

\vspace{-2ex}
\subsection{Hyper-parameter Study: ASR v.s. Poisoning Ratio}
\begin{figure}[htbp]
\centering
\begin{minipage}[t]{0.49\textwidth}
\centering
\includegraphics[width=5.8cm]{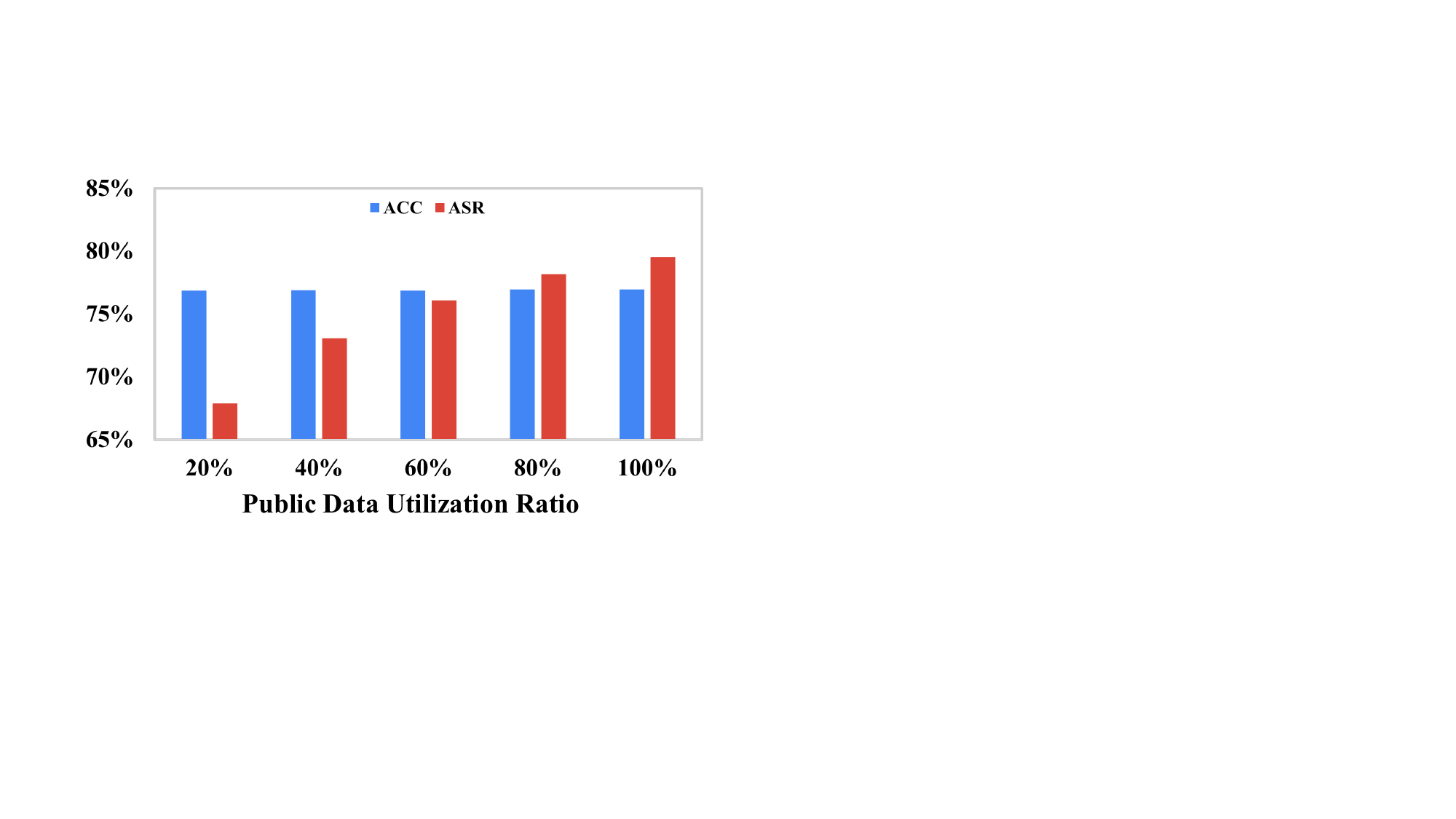}
\caption{\small{Case study of public data utilization.}}\label{fig:case_study}
\end{minipage}
\begin{minipage}[t]{0.48\textwidth}
\centering
\includegraphics[width=5.7cm]{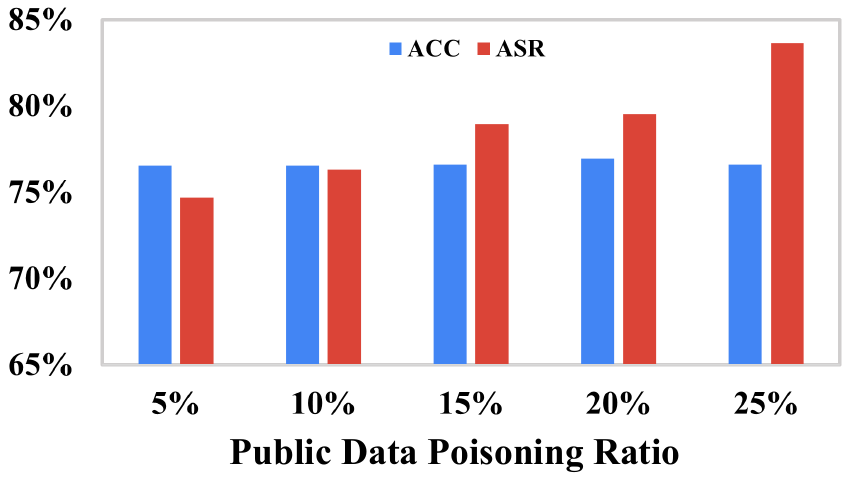}
\caption{\small{Hyperparameter analysis of the poisoning ratio.}}\label{fig:hyper}
\end{minipage}
\vspace{-3ex}
\end{figure}

We further explore the influence of a key hyper-parameter, the poisoning ratio of synthetic data, on the performance of our \ours. 
In our primary experiments on both text and image classification tasks, we set the poisoning ratio to 20\%.
We conduct 4 additional experiments, where we respectively set the poisoning ratio to 5\%, 10\%, 15\%, and 25\%, and the results in terms of ACC and ASR for our proposed attack are shown in Fig.~\ref{fig:hyper}.
These experiments are conducted on the IID CIFAR-10 datasets under the cross-silo FL setting with heterogeneous client model structures.
Similarly, the ACC remains relatively stable despite changes in the poisoning ratio, as the local private training set is untouched. 
As expected, the ASR is positively correlated to the public data poisoning ratio. 
Notably, even at a minimal poisoning ratio of 5\%, our \ours maintains a high level of effectiveness, achieving an ASR of around 75\%.
\begin{table*}[t]
\centering
\resizebox{0.95\textwidth}{!}{
\begin{tabular}{l|l|c|c|c|c|c|c|c|c|c|c|c|c} 
\toprule 
\multicolumn{2}{l|}{\textbf{Setting}} & \multicolumn{6}{c|}{\textbf{Cross-device}} & \multicolumn{6}{c}{\textbf{Cross-silo}}\\\hline
\multicolumn{2}{l|}{\textbf{Approach}} & \multicolumn{2}{c|}{\textbf{Vanilla}} & \multicolumn{2}{c|}{\textbf{CBD}} &\multicolumn{2}{c|}{\ours}  & \multicolumn{2}{c|}{\textbf{Vanilla}} & \multicolumn{2}{c|}{\textbf{CBD}} &\multicolumn{2}{c}{\ours}\\\hline
\multicolumn{2}{l|}{\textbf{Metric}}&Acc&ASR&Acc&ASR&Acc&ASR&Acc&ASR&Acc&ASR&Acc&ASR\\
\hline
\multirow{2}{*}{\textbf{D1}}&\textbf{IID}&
83.70 & 38.24 & 78.81 & 0.22 & 84.59 & 98.92 & 84.49 & 28.33 & 83.46 & 94.68 & 84.24 & 92.61 \\
&\textbf{Non-IID} &
65.16 & 10.22 & 66.76 & 0.01 & 66.63 & 93.37 & 70.18 & 3.37 & 68.12 & 65.13 & 71.17 & 76.94 \\
\hline
\multirow{2}{*}{\textbf{D2}} &\textbf{IID}&
88.83 & 1.18 & 87.67 & 0.34 & 86.67 & 75.79 & 89.33 & 1.18 & 88.60 & 78.83 & 90.13 & 49.91 \\
&\textbf{Non-IID}&
88.33 & 0.05 & 90.99 & 0.48 & 89.00 & 58.57 & 90.67 & 0.89 & 92.33 & 48.54 & 89.67 & 75.82 \\
\hline
\multirow{2}{*}{\textbf{D3}} &\textbf{IID}&
64.43 & 2.66 & 64.47 & 2.72 & 63.21 & 92.89 & 77.52 & 2.84 & 75.92 & 6.85 & 77.27 & 62.57 \\
&\textbf{Non-IID} &
50.58 & 5.62 & 50.51 & 5.42 & 48.24 & 95.16 & 50.46 & 6.98 & 50.82 & 7.83 & 44.92 & 89.71 \\
\bottomrule 
\end{tabular}
}
\caption{Performance (\%) comparison on the text and image classification tasks under the \textbf{homogeneous} setting. D1 is SST-2 dataset, D2 is AG-News, and D3 is CIFAR-10.}
\label{tab:homo}
\vspace{-3ex}
\end{table*}

%% file: section/conclusion.tex
\vspace{-2ex}
\section{Conclusion}


This paper addresses a critical and underexplored aspect of HFL: 
the security vulnerabilities inherent in using FMs for synthetic public dataset generation. 
We unveiled a novel backdoor attack mechanism that can be employed in HFL scenarios without necessitating client compromise or prolonged participation in the FL process. 
Our approach strategically embeds and transfers a backdoor through contaminated public datasets, demonstrating the ability to bypass existing federated backdoor defenses by exhibiting normal client behavior.
Through extensive experiments in various FL settings and on diverse benchmark datasets, we have empirically established the effectiveness and stealth of our proposed attack. 
Our findings reveal a significant security risk in HFL systems using FMs, emphasizing the urgency for developing more robust defense mechanisms in this field.